\documentclass{jps-cp}
\usepackage{txfonts} 

\title{Precise Neutron Lifetime Measurement Using Pulsed Neutron Beams at J-PARC}

\author{N. SUMI$^{1\#}$, K. HIROTA$^2$, G. ICHIKAWA$^3$, T. INO$^3$, Y. IWASHITA$^4$, S. KAJIWARA$^5$, Y. KATO$^5$, M. KITAGUCHI$^6$, K. MISHIMA$^3$, K. MORIKAWA$^7$, T. MOGI$^5$, H. OIDE$^8$, H. OKABE$^7$, H. OTONO$^9$, T. SHIMA$^2$, H. M. SHIMIZU$^6$, Y. SUGISAWA$^{10}$, T. TANABE$^{11}$, S. YAMASHITA$^{11}$, K. YANO$^1$ and T. YOSHIOKA$^9$}

\inst{$^1$Department of Physics, Kyushu University, Fukuoka 819-0395, Japan \\
$^2$Research Center for Nuclear Physics, Osaka University, Ibaraki, Osaka 567-0047, Japan \\
$^3$KEK, High Energy Accelerator Research Organization, Tsukuba 305-0801, Japan \\
$^4$Institute for Chemical Research, Kyoto University, Uji, Kyoto 611-0011, Japan \\
$^5$Department of Physics, The University of Tokyo, Bunkyo-Ku, Tokyo 113-0033, Japan \\
$^6$Kobayashi-Maskawa Institute for the Origin of Particles and the Universe, Nagoya University, Nagoya 464-8602, Japan\\
$^7$Department of Physics, Nagoya University, Nagoya 464-8602, Japan\\
$^8$Department of Physics, Tokyo Institute of Technology, Tokyo 152-8551, Japan\\
$^9$Research Center for Advanced Particle Physics, Kyushu University, Fukuoka 819-0395, Japan\\
$^{10}$Institute of Applied Physics, University of Tsukuba, Tsukuba, Ibaraki 305-8573, Japan\\
$^{11}$International Center for the Elementary Particle Physics, The University of Tokyo, Tokyo 113-0033, Japan}

\email{sumi@epp.phys.kyushu-u.ac.jp}

\recdate{January 10, 2020}

\abst{A neutron decays into a proton, an electron, and an anti-neutrino through the beta-decay process. 
The decay lifetime ($\sim$880 s) is an important parameter in the weak interaction. 
For example, the neutron lifetime is a parameter used to determine the $|V_{ud}|$ parameter of the CKM quark mixing matrix.
The lifetime is also one of the input parameters for the Big Bang Nucleosynthesis, which predicts light element synthesis in the early universe. 
However, experimental measurements of the neutron lifetime today are significantly different (8.4 s or 4.0$\sigma$) depending on the methods. 
One is a bottle method measuring surviving neutron in the neutron storage bottle. 
The other is a beam method measuring neutron beam flux and neutron decay rate in the detector. 
There is a discussion that the discrepancy comes from unconsidered systematic error or undetectable decay mode, such as dark decay.
A new type of beam experiment is performed at the BL05 MLF J-PARC. 
This experiment measured neutron flux and decay rate simultaneously with a time projection chamber using a pulsed neutron beam. 
We will present the world situation of neutron lifetime and the latest results at J-PARC.}

\kword{neutron, lifetime, MLF, J-PARC}

\begin{document}
\maketitle

\section{Introduction}
\begin{figure}[ht]
\begin{center}
\includegraphics[bb= 0 0 696 472, width=80mm]{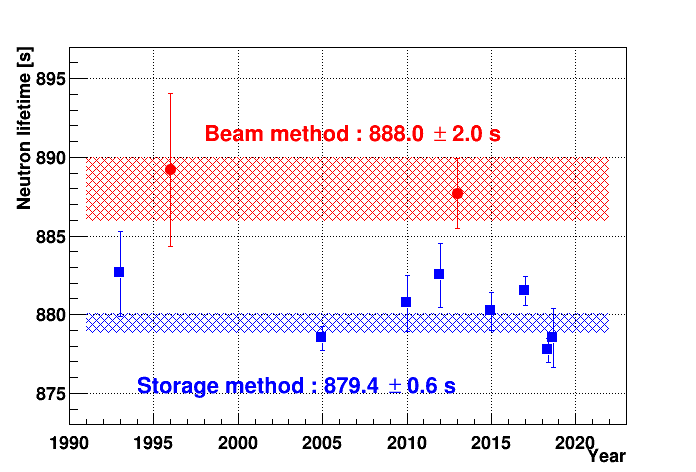}
\caption[The measured neutron lifetime over the publication year]{The measured neutron lifetime over the publication year. There are two types of methods, one is called ``storage method'' and the other is ``beam method''. The discrepancy between these two methods of 8.6 s or 4.1$\sigma$ is called ``neutron lifetime anomaly''.}
\label{fig:lifetimehistory}
\end{center}
\end{figure}

A free neutron decays into a proton, electron, and anti-neutrino with a mean lifetime $\tau_n \sim$ 15 min denoted as,
\begin{equation}
\rm n \rightarrow p + e^- + \overline{\nu}_e .
\end{equation}
Figure~\ref{fig:lifetimehistory} shows the measured neutron lifetime in these twenty years.
There are two types of methods, one is called ``storage method'' and the other is ``beam method''.
The discrepancy between these two methods of 8.6 s or 4.1$\sigma$ is called ``neutron lifetime anomaly''.
Before explaining the measurement methods in detail, the physical significance of $\tau_n$ will be introduced in the next section.

\subsection{Big Bang Nucleosynthesis}
\begin{figure}[ht]
\begin{center}
\includegraphics[ width=80mm]{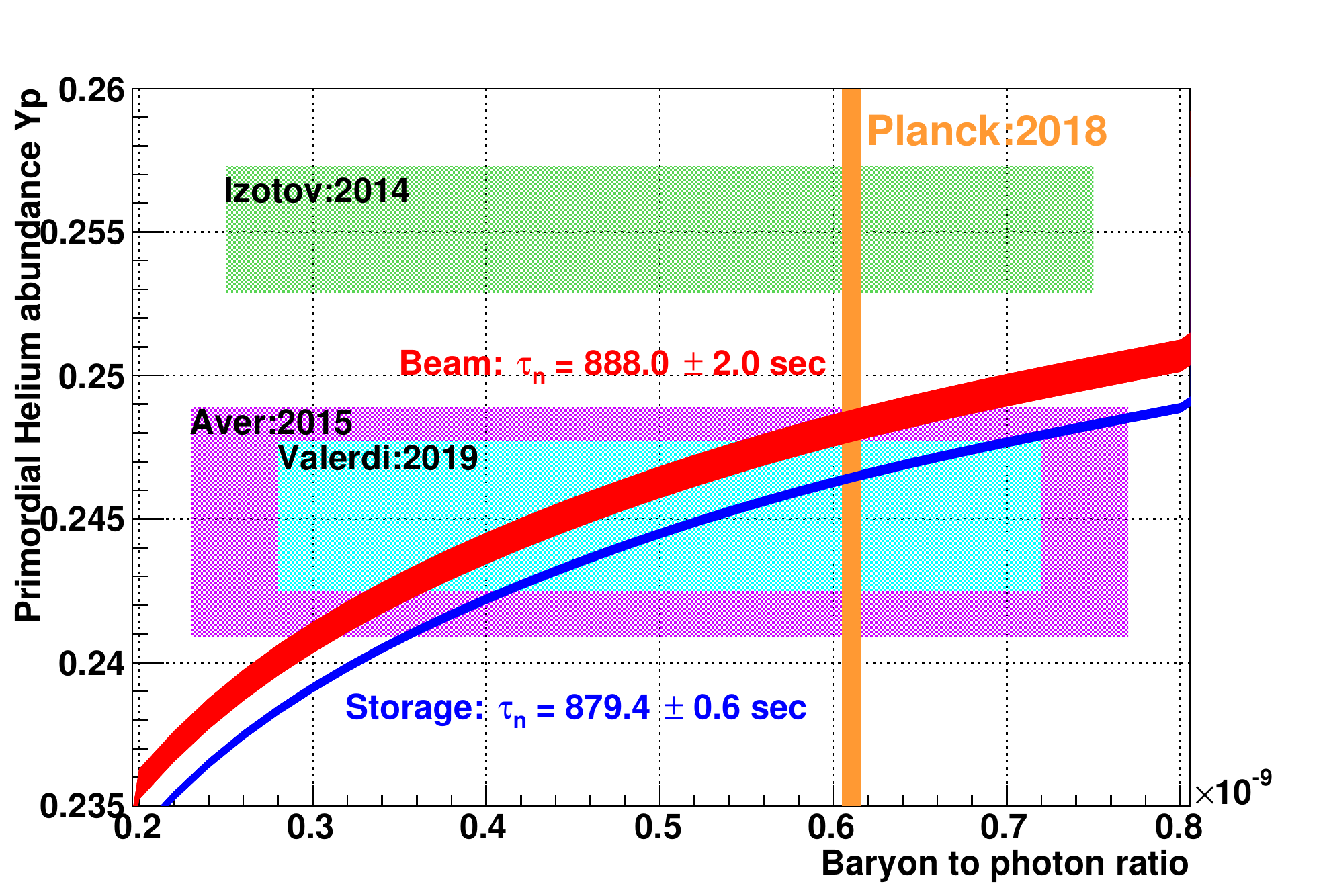}
\caption[The observations and the prediction of helium abundance]{The observations and the prediction of helium abundance $Y_p$. The three filled rectangular regions are observed results of $Y_p$. The vertical region is the baryon to photon ratio $\eta$ and the two curved bands are the prediction of BBN by the two $\tau_n$ results.}
\label{fig:SBBN}
\end{center}
\end{figure}

The Big Bang Nucleosynthesis (BBN) is a theory that estimates the production of the light element in the early universe.
Since the time scale of the BBN is similar to $\tau_n$, the abundance of light nuclei strongly depends on it.
Figure \ref{fig:SBBN} is the observations of the early universe and the prediction of helium abundance $Y_p = \rm {^4He}/(H+{^4He})$.
 The predicted $Y_p$ is the cross point of the band of $\tau_n$ and baryon to photon ratio $\eta$, which is determined by the Planck satellite from the observation of cosmic microwave background (CMB) \cite{collaboration2018planck}.
There are two bands of $\tau_n$ by the measurement methods.
Two observations (Aver:2015\cite{Aver_2015} and Valerdi:2019\cite{Valerdi_2019}) are in good agreement with the prediction, but one observation (Izotov:2014\cite{10.1093/mnras/stu1771}) does not.
Since the observed accuracy of $Y_p$ and $\eta$ is improving year by year, the ambiguity of $\tau_n$ should be resolved.

\subsection{Unitarity of CKM matrix}
\begin{figure}[ht]
\begin{center}
\includegraphics[bb= 0 0 496 472, width=70mm]{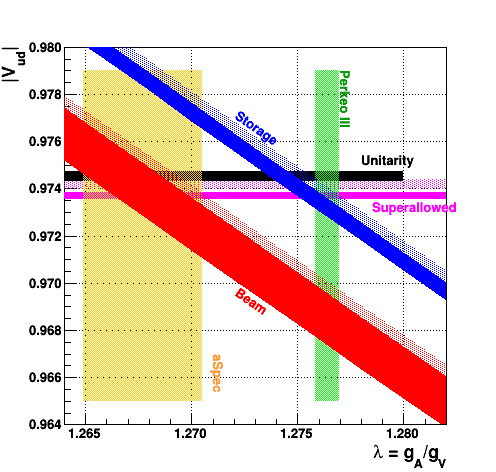}
\caption[$|V_{ud}|$ values along with $\lambda$]{$|V_{ud}|$ values along with $\lambda$. The filled black box indicates $|V_{ud}|$ that satisfies unitarity. The hatched and filled magenta boxes indicate $|V_{ud}|$ by superallowed nuclear decay with an old and a new radiative correction \cite{PhysRevLett.121.241804}, respectively. 
The value obtained from the neutron decay is the cross point of $\tau_n$ and $\lambda$. 
The cross point of $\tau_n$ by storage method and $\lambda$ by Perkeo I\hspace{-1pt}I\hspace{-1pt}I \cite{PhysRevLett.122.242501} and that of beam method and {\it a}SEPCT \cite{beck2019improved} have $|V_{ud}|$ value close to the unitarity. 
Note that, results by the neutron lifetime are also affected by the radiative correction.}
\label{fig:Vud}
\end{center}
\end{figure}

In the standard model of particle physics, the Cabibbo-Kobayashi-Maskawa (CKM) matrix describes the strength of transitions between quarks in weak interactions.
The unitarity check of the CKM matrix gives a strong test of the standard model.
For example, the first row of the matrix gives $\left|V_{u d}\right|^{2}+\left|V_{u s}\right|^{2}+\left|V_{u b}\right|^{2}=0.9994 \pm 0.0005$ \cite{PhysRevD.98.030001}.
The most precise determination of $|V_{ud}|$ comes from the study of superallowed $J^\pi = 0^+ \rightarrow 0^+$ nuclear beta decays, which are pure vector transitions.
The error of $|V_{ud}|$ is dominated by theoretical uncertainties stemming from nuclear Coulomb distortions and radiative corrections. 
A precise determination of $|V_{ud}|$ is also obtained from the measurement of neutron decay as
\begin{equation}
|V_{ud}|^2 = \frac{(4908.7\pm1.9)\; \rm s}{\tau_n(1+3\lambda^2)}.
\end{equation}
The theoretical uncertainties are very small, but the determination is limited by the uncertainties of the ratio of the axial-vector and vector couplings, $\lambda = g_A / g_V$, and $\tau_n$.
Figure~\ref{fig:Vud} shows $|V_{ud}|$ values along with $\lambda$.
The filled black box indicates $|V_{ud}|$ that satisfies unitarity. 
The hatched and filled magenta boxes indicate $|V_{ud}|$ by superallowed nuclear decay with an old and a new radiative correction \cite{PhysRevLett.121.241804}, respectively. 
This value has a slightly smaller value from the unitarity with the correction. 
The value obtained from the neutron decay is the cross point of $\tau_n$ and $\lambda$. 
The cross point of $\tau_n$ by the storage method and $\lambda$ by Perkeo I\hspace{-1pt}I\hspace{-1pt}I \cite{PhysRevLett.122.242501} and that of the beam method and {\it a}SEPCT \cite{beck2019improved} have $|V_{ud}|$ value close to the unitarity.
The value $\lambda$ can be calculated from the QCD lattice gauge theory, but the calculated results cannot reproduce the experimental value \cite{PhysRevD.68.054509}.

\subsection{Neutron dark decay}
To explain the disagreement between ``storage method'' and ``beam method'', Fornal et al. suggest that the neutron cloud decay into unobserved particles by 1\% of the usual $\beta$ decay \cite{PhysRevLett.120.191801}.
In the beam method, experimentalists could not observe unexpected decay mode as neutron decay and the result got longer.
On the other hand, in the storage method, the result would not rely on the decay mode.
They propose that the neutron could decay into dark matter particles with the following decay modes,
\begin{eqnarray}
&n \rightarrow \chi + \gamma	& \;(937.900 \; {\rm MeV} < m_\chi < 938.783 \; {\rm MeV}) \label{eq:darkdecaygamma}\\
&n \rightarrow \chi + e^+ + e^-	& \;(937.900 \; {\rm MeV} < m_\chi < 938.543 \; {\rm MeV}) \label{eq:darkdecayelectron}\\
&n \rightarrow \chi + \phi 		& \;(937.900 \; {\rm MeV} < m_\chi + m_\phi < 939.565 \; \rm MeV) \label{eq:darkdecay},
\end{eqnarray}
where $\phi$ is another dark matter particle.
The mass of dark matter $m_\chi$ and $m_\phi$ are strictly limited by the stability of the proton and nuclei.
After the publication of the neutron dark decay paper, some experiments \cite{PhysRevLett.121.022505, PhysRevLett.122.222503} rejected some decay modes.


\section{Measurement methods} \label{sec:methods}

\subsection{Storage method} \label{sec:storagemethod}
The storage method measures neutron lifetime by storing ultracold neutron (UCN) in the specific bottle.
They counts the number of surviving neutrons $S_1$ and $S_2$ after distinct storing times $t_1$ and $t_2$.
Then, $\tau_n$ is calculated by,
\begin{equation}
\frac{\ln \left(S_{1} / S_{2}\right)}{t_2 - t_1}=\frac{1}{\tau_{n}}+\frac{1}{\tau_{wall}}.
\end{equation}
In this equation, $\tau_{wall}$ is the wall loss effect of the stored neutron.
There are many reasons to lose neutrons from the bottle, e.g. absorption and scattering.
The estimation and correction of the $\tau_{wall}$ is the key point of the storage method.
In this big gravitational trap experiment \cite{serebrov}, the ultracold neutrons were guided and filled into the UCN trap.
After a certain storing time, the survived neutrons were released to the neutron detector below the bottle.
The $\tau_{wall}$ was estimated 1.5\% for the lifetime of the neutron by changing the volume or temperature of the bottle.
This experiment published the result of $\tau_n = \rm 881.5 \pm 0.7 \; (stat) \pm 0.6 \; (syst) \;sec$.

To prevent interaction between neutron and wall material, other experiments \cite{Pattieeaan8895} and \cite{Ezhov} stored neutron by magnetic field potential.
These experiments aligned strong permanent magnets and the neutron, whose spin is parallel to the field, bounce on the potential.
These experiments published the result of $\tau_n = \rm 877.7 \pm 0.7 \; (stat) ^{+0.4}_{-0.2} \; (syst) \; s$ and $\tau_n = \rm 878.3 \pm 1.6 \; (stat) \pm 1.0 \; (syst) \; s$, respectively.

\subsection{Beam method} \label{sec:beammethod}
The beam method measures neutron lifetime by counting the injected neutron and decay product in the beam.
In this penning trap experiment \cite{PhysRevLett.111.222501}, the neutron beam was injected into the volume and the decay proton was stored in the magnetic field and electric field.
The flux of the injected neutron beam was monitored by converted to charged particles at a thin $^6$Li plate via, $\rm ^{6}Li+n \rightarrow \alpha+t$.
Then, these $\alpha$-ray or triton was detected by the surrounding detectors.
After the neutron beam was stopped, the trapped protons were accelerated toward the proton detector along the magnetic field by one side of the electrode voltage was dropped to 0 V.
The neutron lifetime was obtained from the counting ratio of these two detectors.
This experiment published the result of $\tau_n = \rm 887.7 \pm 1.2 \; (stat) \pm 1.9 \; (syst) \; s.$

In another beam method using Time Projection Chamber (TPC) \cite{1989NIMPA.284..120S}, neutron lifetime is obtained from the simultaneous measurement of an electron from $\beta$ decay and $^3$He(n,p)$^3$H reaction.
They chopped neutron beam to short bunch and injected them into TPC.
The neutron lifetime was obtained from,
\begin{equation} \label{eq:neutronlifetime}
\tau_n = \frac{1}{\rho \sigma v} \biggl(\frac{S_{_{\rm He}}/\varepsilon_{_{\rm He}}} {S_\beta/\varepsilon_{\beta}} \biggl)
\end{equation}
where $S_\beta$ and $S_{\rm He}$ are the counting numbers of $\beta$ decay and $^3$He(n,p)$^3$H reaction, $\varepsilon_{\beta}, \varepsilon_{\rm He}$ are the detection efficiency of them, $v$ is velocity of the neutron, $\rho$ and $\sigma$ are the number density and absorption cross section of $^3$He.
This experiment published the result of $\tau_n = \rm 878 \pm 27 \; (stat) \pm 14 \; (syst) \; s.$
Its accuracy was limited by the statistics and the background for the $\beta$ decay signals.

\section{Neutron lifetime measurement at J-PARC} 
\begin{figure}[ht]
\begin{center}
\includegraphics[ width=120mm]{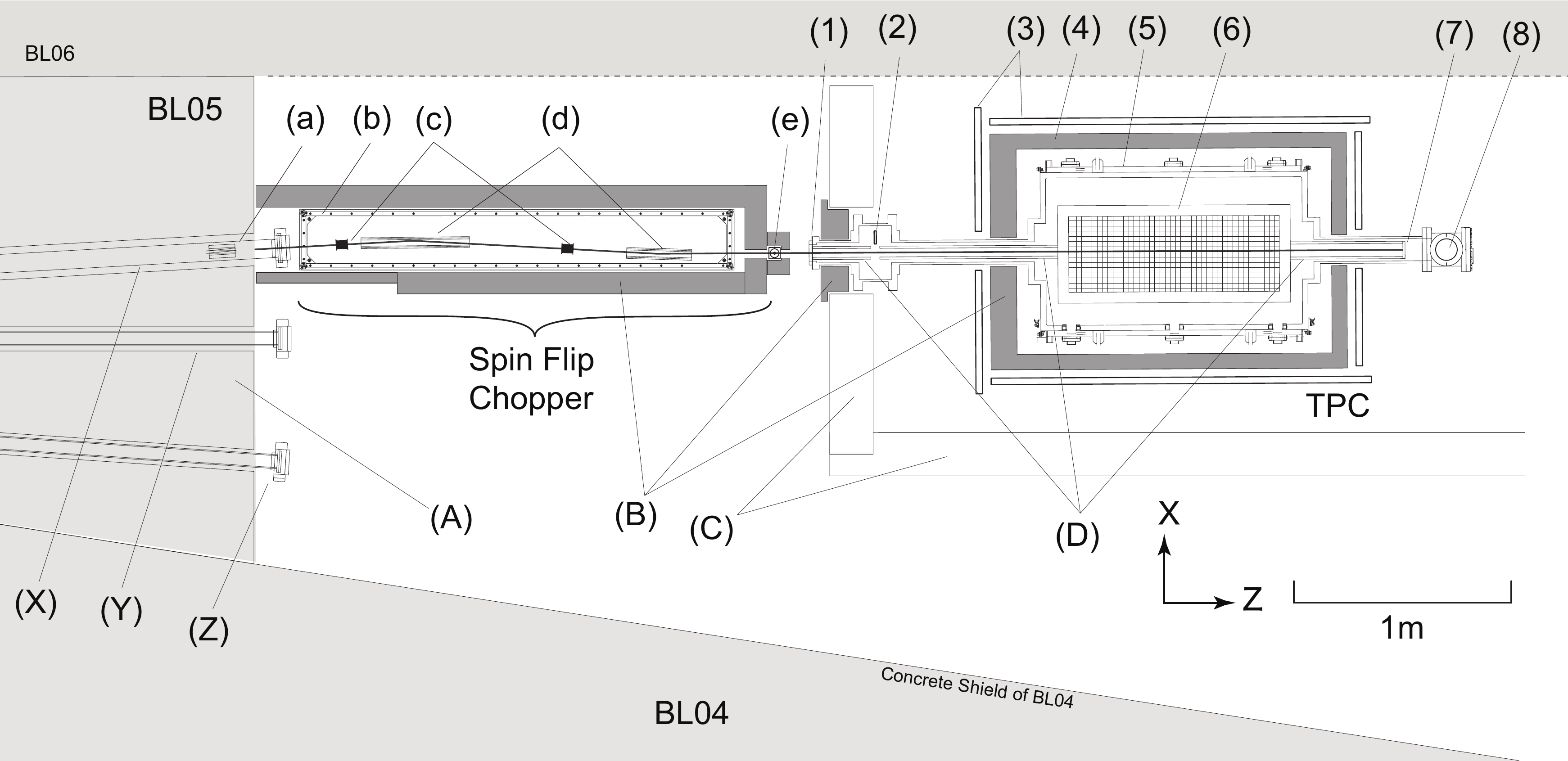}
\caption[Schematic view of the neutron lifetime measurement at BL05.]{Schematic view of the neutron lifetime measurement at BL05. (X) Polarized (Y) High intensity (Z) Low divergence beam branch (A) Concrete shield (B) Lead shield (C) Iron shield (D) $^6$LiF neutron duct (a) Collimator (b) Guide coil (c) Spin Flipper (d) Magnetic super mirror (e) Neutron beam monitor (1) Zr window (2) $^6$LiF neutron shutter (3) Cosmic veto counter (4) Lead shield (5) Vacuum chamber (6) Time Projection Chamber (7) Beam dump (8) Vacuum pump}
\label{fig:bl05}
\end{center}
\end{figure}

Figure~\ref{fig:bl05} is a schematic view of the neutron lifetime measurement at BL05 \cite{BL05} in the Materials and Life Science Experimental Facility (MLF), Japan Proton Accelerator Research Complex (J-PARC).
The neutron beam is chopped at the spin flip chopper (SFC) to make a short bunch and injected it to the TPC.
The neutron shutter is installed at the upstream of the TPC to control the neutron bunch.
The TPC counts the $\beta$ decay and $^3$He absorption signals.
The rest of the bunch is absorbed in the beam dump.

The neutron beam data have been acquired since 2014.
In this paper, the data from 2014 to 2016 are used to analyze.
These acquired data are summarized in table \ref{tab:aquireddata}.

\begin{table}[ht]
\begin{center}
\begin{tabular}{cccc} \hline
Gas	&Date		&MLF power [kW]	&Beam time [hour]	\\ \hline \hline
I	&May 2014	&300				&35.3			\\
II	&April 2015	&500				&15.8			\\
III	&April 2016	&200				&17.5			\\
IV	&April 2016	&200				&72.7			\\
V	&May 2016	&200				&69.4			\\
VI	&June 2016	&200				&71.1			\\ \hline
\end{tabular}
\caption[Acquired data from 2014 to 2016]{Acquired data from 2014 to 2016. One data set is corresponding to one gas fill. MLF power denotes J-PARC proton beam power to which the neutron beam power is proportional.}
\label{tab:aquireddata}
\end{center}
\end{table}

The number of two signals $S_\beta, S_{\rm He}$ are extracted from the acquired data using signal selection cut.
The time-independent backgrounds are subtracted using the time of fight method and the neutron shutter open and closed data.
The cut efficiencies $\varepsilon_\beta, \varepsilon_{\rm He}$, and the number of remaining backgrounds are estimated using Monte Carlo simulation.
Then, the neutron lifetime $\tau_n$ can be calculated by equation~\ref{eq:neutronlifetime}.

\subsection{Signal selection} \label{sec:signalselection}
The selection for the neutron $\beta$ decay signal is the following five cuts.
The first one is ``time of flight cut'' which requires that event trigger time is in the neutron bunch completely inside the TPC ($-380$ mm$<z<380$ mm).
The second is ``drift length cut'' which requires that drift length, or $y$ length, is smaller than half of the TPC ($<$190 mm).
The third is ``distance from beam axis cut'' which requires that the edge of the track on the beam axis within $\pm$48 mm.
The fourth is ``point like cut'' which requires that the range of the track is greater than 100 mm or deposit energy is greater than 5 keV to eliminate CO$_2$ recoil point-like event.
The last is ``high energy cut'' which requires that the energy on a low gain wire is smaller than 25 keV for all wires to eliminate $^3$He absorption.

The selection for the neutron $^3$He absorption signal is following two cuts.
The first one is ``time of flight cut'' which is the same as $\beta$ decay signal.
The second one is an inversion of ``high energy cut'' which requires that any of the low gain wire exceeds 25 keV.

\subsection{Background subtraction}
\begin{figure}[ht]
\begin{center}
\includegraphics[ width=80mm]{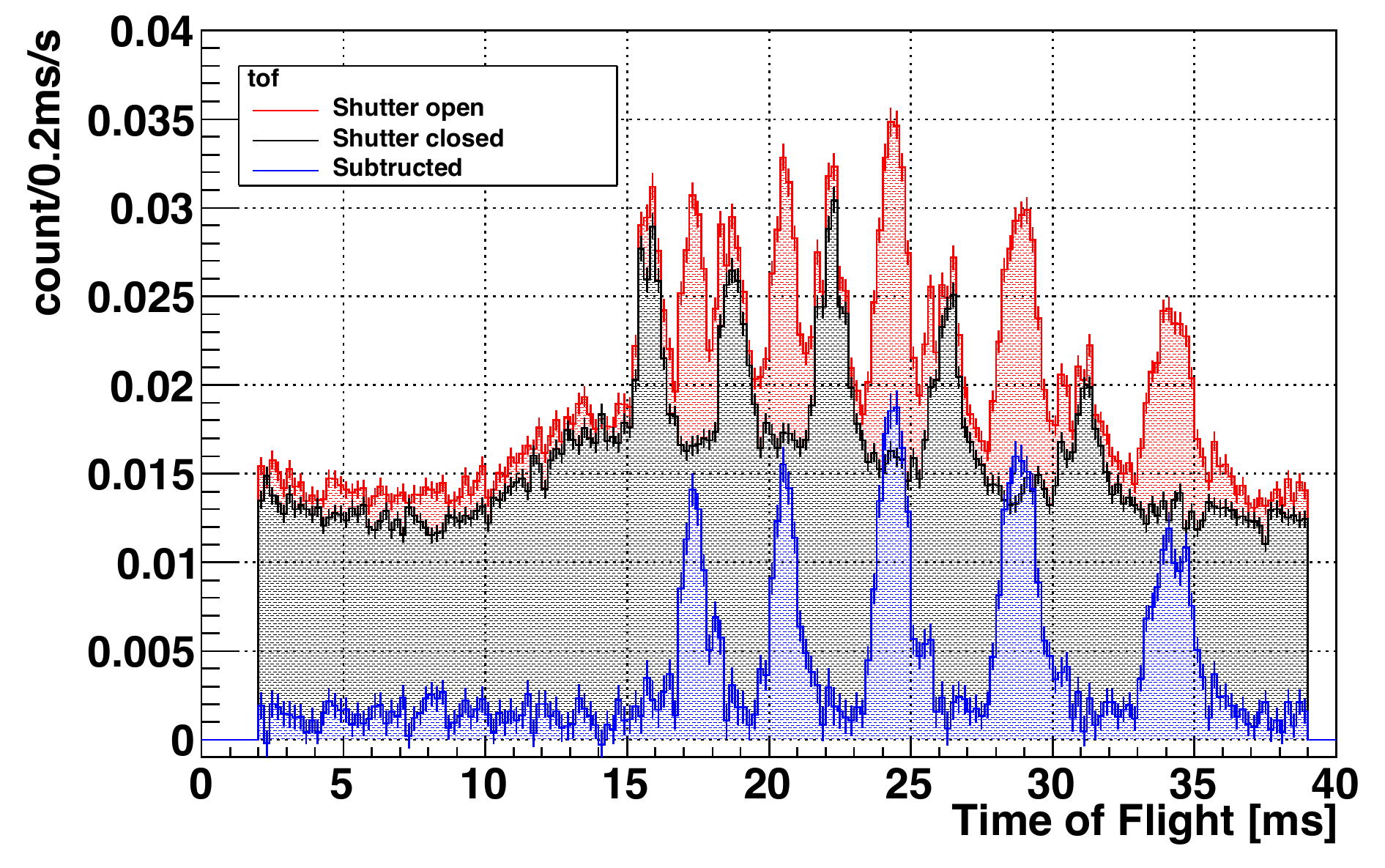}
\caption[Time of flight drawn by acquired data Gas II]{Time of flight drawn by acquired data Gas II. The red and black filled areas are the neutron shutter open and closed state and the blue area is a subtraction of them. There are five clear peaks on the small flat background in the subtraction spectrum.}
\label{fig:tof}
\end{center}
\end{figure}

The time-independent background is subtracted using time of flight method and neutron shutter open and closed data.
The $\beta$ decay and neutron-induced background emerge only while the neutron bunch in the TPC, defined as ``fiducial time''. 
Ahead of it, upstream $\gamma$-ray coming from SFC generates background peak and more background comes from the mercury target at the time of flight $T=0$. 
The quiet time between them is defined as ``sideband time''.
Figure~\ref{fig:tof} is the time of flight drawn by acquired data.
The red and black filled areas are the neutron shutter open and closed state and the blue area is a subtraction of them.
There are five clear peaks on the small flat background in the subtraction spectrum.
The flat background is TPC internal wall activation of $^{20}$F ($\tau_{1/2}$ = 11.2 s) and $^8$Li ($\tau_{1/2}$ = 840 ms).
Since they have a longer lifetime than the MLF beam cycle of 40 ms, they are regarded as time constant background.
However, their lifetimes are shorter than an interval of the shutter open and closed, they disappear at closed operation.

\subsection{Background estimation}
The remaining backgrounds for $S_\beta$ are estimated using Monte Carlo simulation.
In the CO$_2$ capture reaction$\rm ^{12}C + n \rightarrow \,^{13}C \; (1 keV) + \gamma \; (4945 keV)$, the recoiled $^{13}$C deposits small energy on the beam axis, but 99.9\% events are eliminated by ``point like cut''.
However, if the $\gamma$-ray scatters electron in TPC, it turns to the background.
The scattered neutron $\beta$ decay is treated as background because they have an unpredictable track for each decay point.
The wall capture $\gamma$-ray is the dominant source of the remaining background.
The detector wall captures scattered neutron and emits $\gamma$-ray.
If the $\gamma$-ray scatter an electron, it turns to the background.

\begin{figure}[ht]
\begin{center}
\includegraphics[ width=70mm]{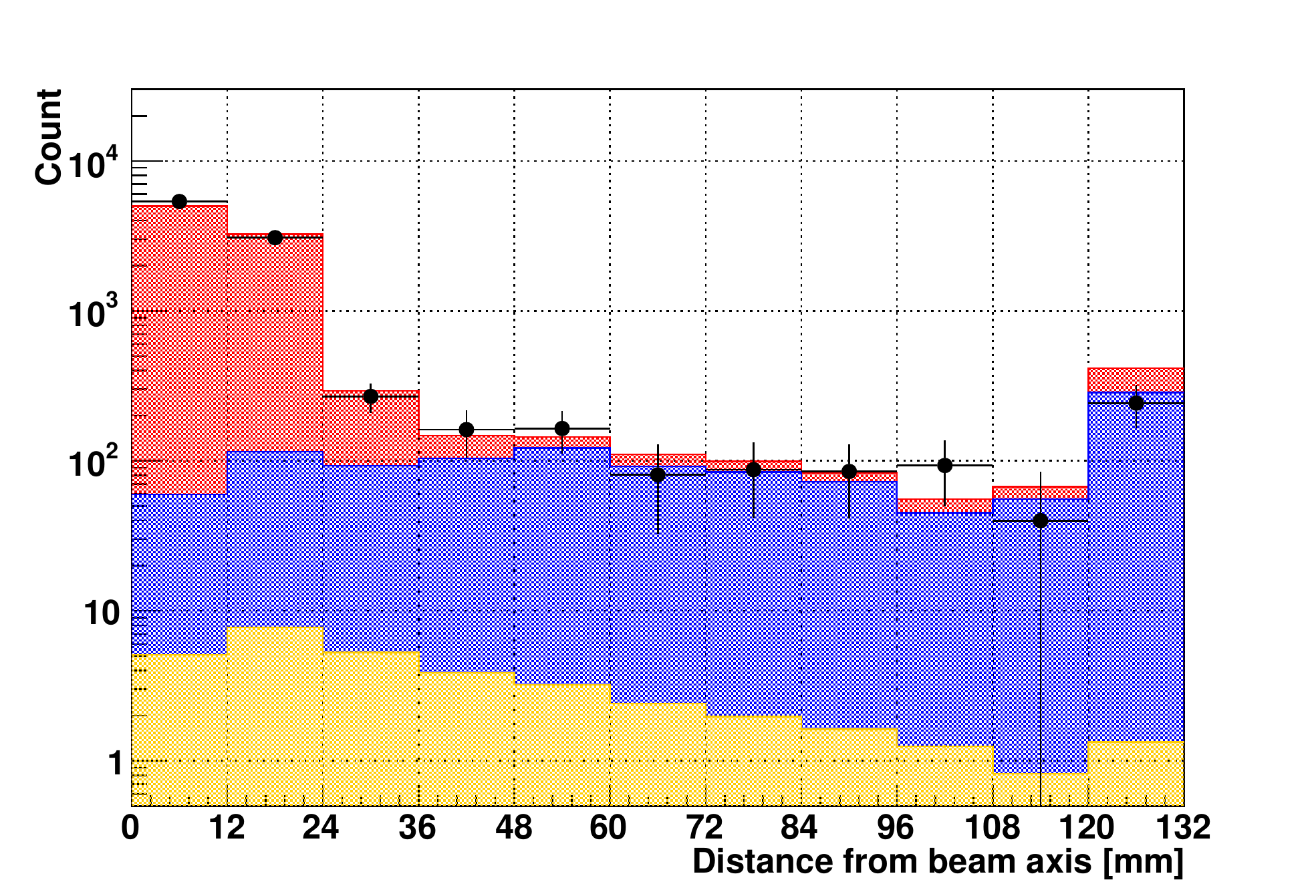}
\caption[Background contamination for $S_\beta$ over distance from beam axis]{Background contamination for $S_\beta$ over distance from beam axis. The distance = 0 mm represents track edge on beam axis, the distance $>$ 120 mm represents track edge on the wall. The black point indicate experimental data. The stacking histograms are simulation data, scattered $\beta$ decay (orange), wall capture $\gamma$-ray (blue), and $\beta$ decay (red) from the bottom.}
\label{fig:backgroundcontamination}
\end{center}
\end{figure}

Figure~\ref{fig:backgroundcontamination} shows estimation of the background contamination over distance from beam axis. 
The distance = 0 mm represents track edge on the beam axis, the distance $>$ 120 mm represents track edge on the wall.
The black point indicates experimental data. 
The stacking histograms are simulation data, scattered $\beta$ decay (orange), wall capture $\gamma$-ray (blue), and $\beta$ decay (red) from the bottom.
The simulation data were scaled to the off-axis control region.
The estimated contamination is 463 $\pm$ 154 event in total.
The contribution of the CO$_2$ capture reaction is negligible.

The backgrounds for $S_{\rm He}$ are following two absorption on the beam axis,
\begin{eqnarray}
\rm ^{14}N + n &\rightarrow& \rm ^{14}C + p + 626keV, \\
\rm ^{17}O+n &\rightarrow& \rm ^{14}C+\alpha+1818 keV.
\end{eqnarray}
Although they have a different energy from $^3$He absorption of 764 keV, the gain of Multi Wire Proportional Chamber (MWPC) is saturated at their energy region in high gain operation.
Therefore, low gain operation data were taken once a day to measure contamination of them.
The source of nitrogen is outgassing in the vacuum chamber and its rate was estimated at 0.4 Pa/day and corrected its contribution.
The source of oxygen is CO$_2$ as the quencher gas of 15 kPa.
Unlike $^{14}$N, fluctuation by outgassing is negligible, thus its contribution was calculated by the cross section and natural abundance as a time constant and corrected 0.50\% for $^3$He of 100 mPa.
Besides, $^3$He absorption of the scattered neutron is treated as background same as scattered neutron $\beta$ decay.
The contribution of such an event was evaluated by 0.30\% by the amount of off-axis $^3$He absorption.

\section{Result and uncertainty}
Table~\ref{tab:result} is results and uncertainties of the all values in equation~\ref{eq:neutronlifetime} for the Gas II data.
The number of $\beta$ decay signal $S_\beta$ has the largest uncertainty in the values due to background contamination.
The efficiency of $\beta$ decay signal $\varepsilon_\beta$ has the next largest uncertainty.
The number density of $^3$He gas will be improved to 0.1\% with the updated injection method.
Another experiment is planned to improve the accuracy of the cross section $\sigma_0$ of $^3$He.

\begin{table}[ht]
\begin{center}
\small
\begin{tabular}{crrrr}\hline
Value				&Result						&Correction		&Uncertainty	&Note	\\ \hline
$\rho$				&(2.08$\pm$0.01) $10^{19}$/m$^3$	&0				&0.5\%	&Improved by $^3$He gas injection\\
$\sigma_0 v_0$		&5333$\pm$7 barn $\times$ 2200 m/s	&0			&0.13\%	&Requires other measurement\\
$S_{\rm He}$			&202993 $\pm$ 480 				&2672 $\pm$ 351 	&0.3\%	&Statistical error\\
$S_\beta$				&8868 $\pm$ 151 				&463 $\pm$ 154 	&2.6\%	&Background contamination\\
$\varepsilon_{\rm He}$	&(100$-$0.014)\%				&0\%			&0.014\%	&Enough accuracy\\
$\varepsilon_\beta$		&(94.5$\pm$0.7)\%				&(+5.5$\pm$0.7)\%	&0.7\%	&Simulation uncertainty\\ \hline
\end{tabular}
\caption{The results and uncertainties for the Gas II.}
\label{tab:result}
\end{center}
\end{table}

The combined result of all Gas I - VI is,
\begin{equation}
\tau_n = 896 \pm 10 \; {\rm (stat)} \; ^{+14}_{-10} {\rm \; (syst) \; s}.
\end{equation}
This result has still large uncertainty to compare the other results, but it is consistent with the beam method and storage method.
Since this method is independent of the other methods, the improved result gives a hint to discuss the disagreement.

In addition, we published an updated result,
\begin{equation}
\tau_n = 898 \pm 10 \; {\rm (stat)} \; ^{+15}_{-18} {\rm \; (syst) \; s},
\end{equation}
by reconsidering the systematic uncertainty \cite{BL05newresult}.

\section*{Acknowledgements}
This research was supported by JSPS KAKENHI Grant Number 19GS0210 and JP16H02194. 
The neutron experiment at the Materials and Life Science Experimental Facility of the J-PARC 
was performed under a user program (Proposal No. 2015A0316, 2014B0271, 2014A0244, 2012B0219, and 2012A0075) and 
S-type project of KEK (Proposal No. 2014S03).

\end{document}